\documentstyle[aps,prd,epsfig,twocolumn,floats]{revtex}
\def\Dslash{D\hspace*{-6.5pt}\raisebox{1pt}{/}\hspace*{1pt}}
\def\pslash{\partial\hspace*{-5pt}\raisebox{1pt}{/}\hspace*{.5pt}}
\def\Aslash{A\hspace*{-5.5pt}\raisebox{1pt}{/}\hspace*{1pt}}
\begin{document}
\draft
\wideabs{
\title{Microscopic universality with dynamical fermions}
\author{M.E.~Berbenni-Bitsch$^1$, S.~Meyer$^1$, and T.~Wettig$^2$}
\address{$^1$Fachbereich Physik -- Theoretische Physik, Universit\"at
  Kaiserslautern, D-67663 Kaiserslautern, Germany\\
  $^2$Institut f\"ur Theoretische Physik, Technische Universit\"at
  M\"unchen, D-85747 Garching, Germany} 
\date{\today}  
\maketitle
\begin{abstract}
  It has recently been demonstrated in quenched lattice simulations
  that the distribution of the low-lying eigenvalues of the QCD Dirac
  operator is universal and described by random-matrix theory.  We
  present first evidence that this universality continues to hold in
  the presence of dynamical quarks.  Data from a lattice simulation
  with gauge group SU(2) and dynamical staggered fermions are compared
  to the predictions of the chiral symplectic ensemble of
  random-matrix theory with massive dynamical quarks.  Good agreement
  is found in this exploratory study.  We also discuss implications of
  our results.
\end{abstract}
\pacs{PACS numbers: 11.15.Ha, 05.45.+b, 11.30.Rd, 12.38.Gc}}

\narrowtext

It was conjectured a few years ago \cite{Leut92,Shur93} that the
microscopic spectral properties of the QCD Dirac operator, in
particular the distribution of the low-lying eigenvalues, are
universal and can be obtained in effective theories that are much
simpler than QCD.  This conjecture rests on the observation
\cite{Gass87} that in the range $1/\Lambda\ll V^{1/4}\ll 1/m_\pi$,
where $\Lambda$ is a typical hadronic scale, $V$ is the space-time
volume, and $m_\pi$ is the pion mass, the mass dependence of the
finite-volume partition function of QCD is completely determined by
global symmetries.  Thus, the spectral properties of the Dirac
operator on the ``microscopic'' scale $\sim 1/(V\Sigma)$, where
$\Sigma$ is the absolute value of the chiral condensate, can be
computed in effective theories with only the global symmetries as
input, such as the framework of effective Lagrangians \cite{Leut92} or
chiral random-matrix theory (RMT) \cite{Shur93}.

The global spectral density of the Euclidean Dirac operator
$\Dslash=\pslash+ig\Aslash$ is given by
$\rho(\lambda)=\langle\sum_n\delta(\lambda-\lambda_n)\rangle_A$, where
the $\lambda_n$ are the eigenvalues of $i\Dslash$ and the average is
over all gauge field configurations $A$ weighted by $\exp(-S_{\rm
  QCD})$.  The Banks-Casher formula, $\Sigma=\pi\rho(0)/V$
\cite{Bank80}, relates the spectral density at zero virtuality to the
chiral condensate.  If chiral symmetry is spontaneously broken, this
relation implies that the spacing of the low-lying eigenvalues is
$\sim 1/(V\Sigma)$, rather than $\sim 1/V^{1/4}$ as in the case of the
non-interacting Dirac operator.  Thus, the distribution of the
low-lying eigenvalues is of great interest for a better understanding
of the phenomenon of spontaneous chiral symmetry breaking.

It has recently been shown that the microscopic spectral properties of
the staggered lattice Dirac operator in quenched SU(2) are indeed
universal and described by chiral RMT \cite{Berb98,Ma98}.  However,
for a better understanding of hadronic properties it is important to
go beyond the quenched approximation.  One of the problems that arises
in unquenched lattice simulations is that one cannot go to arbitrarily
small quark masses.  An important point is how small the masses of the
dynamical quarks have to be so that they are really dynamical, i.e.,
lead to results that are different from those obtained in the quenched
approximation.  In this context, the main question that will be
addressed in this work is the following.  What is the effect of light
dynamical quarks with masses of order $\sim 1/(V\Sigma)$ on the
low-lying spectrum of the Dirac operator, and can this effect be
described by results obtained in chiral RMT?  The answer to these
questions together with the analytical information available from RMT
leads to a better understanding of the mass and energy scales in the
problem and has important technical implications with regard to
extrapolations to various limits (chiral, thermodynamic, continuum)
that are difficult to take on the lattice.  Moreover, analytical
knowledge of the distribution of the smallest eigenvalues may be
interesting from an algorithmic point of view.

We will mainly be concerned with the distribution of the smallest
positive eigenvalue, $P(\lambda_{\rm min})$, and with the microscopic
spectral density, $\rho_s(z)$, defined by \cite{Shur93}
\begin{equation}
  \label{eq2}
  \rho_s(z)=\lim_{V\to\infty}\frac{1}{V\Sigma}\:
  \rho\left(\frac{z}{V\Sigma}\right)\:.
\end{equation}
This definition amounts to a magnification of the region of low-lying
eigenvalues by a factor of $V\Sigma$ and leads to the resolution of
individual eigenvalues.  The claim is that the quantities
$P(\lambda_{\rm min})$ and $\rho_s(z)$ are universal (see
Ref.~\cite{Berb98} for a summary of existing evidence) and can be
computed exactly in an effective theory.  We will concentrate on
chiral RMT although identical results could also be obtained from the
finite-volume partition function computed from an effective Lagrangian
\cite{Smil95,Damg97}.

Let us recall how the random-matrix model is constructed.  In a chiral
basis, the weight function of full QCD with $N_f$ flavors can be
written as 
\begin{equation}
  \label{eq5}
  e^{-S_{\rm QCD}}=e^{-S_{\rm gl}}\prod_{f=1}^{N_f}\det\left[
    \matrix{m_f&iT\cr iT^\dagger&m_f}\right]\:,
\end{equation}
where $S_{\rm gl}$ is the gluonic action, $T$ is a matrix representing
$i\Dslash$, and the $m_f$ are the masses of the dynamical quarks.  We
now replace $T$ by a random matrix $W$ with $N$ rows and $N+\nu$
columns ($|\nu|\ll N$).  The dimensionless space-time volume can be
identified with $2N$, and $\nu$ plays the role of the topological
charge.  The symmetry properties of $W$ depend on the gauge group and
on the representation of the fermions ~\cite{Verb94a}.  We will be
concerned with staggered fermions in SU(2) for which the elements of
$W$ are real quaternions.  This is the chiral symplectic ensemble
(chSE).  The average over gauge field configurations is replaced by an
average over random matrices, i.e., the factor $e^{-S_{\rm gl}}$ is
replaced by a convenient distribution of the matrix $W$, often a
simple Gaussian, but the results are insensitive to this choice
\cite{Nish96,Sene98,Nish98}.  For the present work, the second factor
in Eq.~(\ref{eq5}) is more interesting.  In terms of the eigenvalues
of $W$, it reads for the chSE \cite{Verb94a}
\begin{equation}
  \label{eq6}
  \prod_{f=1}^{N_f} \det(WW^\dagger+m_f^2)\propto
  |\Delta(\lambda^2)|^4\prod_n\lambda_n^{4\nu+3}
  \prod_{f=1}^{N_f}(\lambda_n^2+m_f^2)
\end{equation}
in a sector of topological charge $\nu$, where $\Delta$ is the
Vandermonde determinant.  Thus, in the random-matrix model the fermion
determinants can be taken into account without further assumptions.
From Eq.~(\ref{eq6}) it is intuitively clear that the presence of the
fermion determinants will affect the microscopic spectral quantities
only if the $m_f$ are on the order of the smallest eigenvalue or, in
other words, on the order of the mean level spacing near zero
\cite{Shur93,Jurk96}.  Thus, we require $m_f\sim 1/(V\Sigma)$ to
observe an effect on $P(\lambda_{\rm min})$ and $\rho_s(z)$.  For
quark masses much larger than this, we should simply obtain agreement
with the RMT-results computed in the quenched approximation.  This was
already observed in Ref.~\cite{Verb96} where the lattice data of
Ref.~\cite{Chan95} were analyzed.

RMT-predictions for the microscopic spectral quantities in the
presence of massive dynamical quarks have recently been computed for
the chiral unitary ensemble \cite{Damg97a,Wilk97} (see also
\cite{Jurk96}) and the unitary ensemble \cite{Damg97c}.  However, we
want to compare lattice data with random-matrix predictions for the
chSE.  In this case, closed analytical expressions are at present only
known in the chiral limit \cite{Forr93,Naga95,Naga98}.  Analytical
work for the massive case is in progress, but in the meantime we have
computed the RMT-predictions for the chSE with massive quarks
numerically using the methods of Ref.~\cite{Naga95}.  Essentially, one
has to do an iterative computation of skew-orthogonal polynomials
which obey orthogonality relations determined by a weight function
involving the fermion determinants.  To avoid cancellation problems,
we have used a multi-precision package \cite{Bail94}.  The calculation
was done with matrices $W$ of finite dimension $N$ up to $N=100$.  The
remaining finite-$N$ effects can be estimated from a calculation in
the chiral limit and are on the order of 1\%.

We now summarize some RMT-results for the chSE in the chiral limit
which we will need in the following.  With $\alpha=N_f+2|\nu|$, we have
\cite{Naga95,Berb97b,Ma98}
\begin{eqnarray}
  \label{eq7}
  \rho_s(z)&=&z\left[J_{\alpha}^2(2z)-J_{\alpha+1}(2z)J_{\alpha-1}(2z)
  \right]\nonumber\\
  &&-\frac{1}{2}J_{\alpha}(2z)\int_0^{2z}dtJ_{\alpha}(t) \:,
\end{eqnarray}
where $J$ denotes the Bessel function.  Rescaling $\lambda_{\rm min}$
by $V\Sigma$ as in Eq.~(\ref{eq2}), we have for $N_f=\nu=0$
\cite{Forr93}
\begin{equation}
  \label{eq8}
  P(\lambda_{\rm min})=\sqrt{\frac{\pi}{2}}\lambda_{\rm min}^{3/2}
  I_{3/2}(\lambda_{\rm min})\,e^{-\frac{1}{2}\lambda_{\rm min}^2}\:,
\end{equation}
where $I$ is the modified Bessel function.  Further analytical results
for $P(\lambda_{\rm min})$ are known if $\alpha$ is odd, i.e., for odd
$N_f$.  For example, we have for $N_f=1$ and $\nu=0$ \cite{Naga98}
\begin{equation}
  \label{eq9}
  P(\lambda_{\rm min})=\frac12\lambda_{\rm min}
  [2I_2(2\lambda_{\rm min})-I_0(2\lambda_{\rm min})+1]\,
  e^{-\frac{1}{2}\lambda_{\rm min}^2}\:.  
\end{equation}
We shall also need $P(\lambda_{\rm min})$ for $N_f=2$ and 4 which in the
absence of closed analytical expressions can perhaps most easily be
obtained from results computed by Kaneko \cite{Kane93}.  The essential
ingredients are the zonal polynomials at the symmetric point.  We
obtain after some algebra
\begin{equation}
  \label{eq10}
  P(\lambda_{\rm min})=\frac{2}{(\alpha+1)!(\alpha+3)!}
  \lambda_{\rm min}^{2\alpha+3}e^{-\frac{1}{2}\lambda_{\rm min}^2}
  T(\lambda_{\rm min}^2)\:,
\end{equation}
where $T(x)=1+\sum_{d=1}^\infty a_d\,x^d$ with
\begin{eqnarray}
  \label{eq11}
  a_d&=&
  \sum_{{|\kappa|=d}\atop{l(\kappa)\le\alpha+1}}
  \prod_{(i,j)\in\kappa}
  \frac{\alpha\!+\!2j\!-\!i}{\alpha\!+\!2j\!-\!i\!+\!4}\nonumber\\
  &&  \;\times\,\frac{1}{[\kappa_j'\!-\!i\!+\!2(\kappa_i\!-\!j)\!+\!1]
    [\kappa_j'\!-\!i\!+\!2(\kappa_i\!-\!j)\!+\!2]}\:.
\end{eqnarray}
Here, $\kappa$ denotes a partition of the integer $d$, $l(\kappa)$ its
length, $|\kappa|$ its weight, and $\kappa'$ the conjugate partition.
In Eq.~(\ref{eq11}), a partition $\kappa$ is identified with its
diagram, $\kappa=\{s=(i,j); 1\le i\le l(\kappa),1\le j\le\kappa_i\}$.
The Taylor series for $T(x)$ is rapidly convergent, and the curves for
$N_f=2$ and 4 can easily be computed to any desired accuracy.

The RMT-results for $P(\lambda_{\rm min})$ and $\rho_s(z)$ are
sensitive to the topological charge $\nu$.  Thus, they are only
universal in a given sector of topological charge.  In the continuum
limit, the overall result for a given quantity is a weighted average
over these sectors.  RMT can predict the quantity for definite $\nu$
but cannot predict the corresponding weights.  In Ref.~\cite{Berb98}
it was found that for values of $\beta=4/g^2$ up to 2.4 the quenched
SU(2) lattice data were consistent with $\nu=0$.
Here, we only consider values of $\beta$ in the strong-coupling
region, thus everything should be described by the RMT-results for
$\nu=0$.  We will not address the question of topology in the
following.

Our lattice simulations were performed with gauge group SU(2) and
staggered fermions on an $8^4$ lattice.  The boundary conditions were
periodic for the gauge fields and periodic in space and anti-periodic
in Euclidean time for the fermions.  A hybrid Monte Carlo algorithm
\cite{Duan87,Meye90} was used to generate a large number of
independent configurations.  For the diagonalization of $-\Dslash^2$
we have employed the Cullum-Willoughby version of the Lanczos
algorithm \cite{Cull81}.  The complete spectra where checked against
the analytical sum rule ${\rm tr}(-\Dslash^2)=V$ for the distinct
eigenvalues of $-\Dslash^2$ \cite{Kalk95}.  All physical quantities
will be quoted in units of the lattice spacing $a$, in contrast to
Ref.~\cite{Berb98} where the unit was $2a$.

The RMT-results contain one single parameter $V\Sigma$ which is
determined by the lattice data via the Banks-Casher relation,
$V\Sigma=\pi\rho(0)$.  (Note that $\rho(\lambda)$ is normalized to
$V$.)  Thus, the random-matrix predictions are parameter-free.  The
simulation parameters and observables are summarized in
Table~\ref{table1}.  We also give values for
$\langle\bar\psi\psi\rangle(m)= (m/V)
\sum_{\lambda_n>0}(\lambda_n^2+m^2)^{-1}$.  Note that we have
$\mu=mV\Sigma\sim{\cal O}(1)$ as desired, where by ${\cal O}(1)$ we
mean much smaller than $8^4=4096$.
\begin{table}[-t]
  \caption{Simulation parameters and observables of our lattice
    calculations on an $8^4$ lattice.  $\langle\bar\psi\psi\rangle(m)$
    and $\Sigma$ are given in units of $a^{-3}$.} 
  \vspace*{2mm}
  \begin{tabular}{cclrllc}
    $\beta$ & $\tilde N_f$ & \multicolumn{1}{c}{$ma$} & 
    \multicolumn{1}{c}{conf.} & 
    \multicolumn{1}{c}{$\langle\bar\psi\psi\rangle(m)$} & 
    \multicolumn{1}{c}{$\Sigma$} & $mV\Sigma$ \\[0.5mm]\tableline\\[-3mm]
    1.8 & 4 & 0.015 & 1799 & 0.1517(14) & 0.1250(46) & 7.68 \\
    1.3 & 8 & 0.0075 & 1023 & 0.2669(12) & 0.2846(38) & 8.74 \\
    1.3 & 8 & 0.0055 & 1042 & 0.2470(11) & 0.2800(45) & 6.31 \\
    1.3 & 8 & 0.00375 & 703 & 0.2073(22) & 0.2772(42) & 4.26
  \end{tabular}
\label{table1}
\end{table}

\begin{figure}
  \centerline{\epsfig{figure=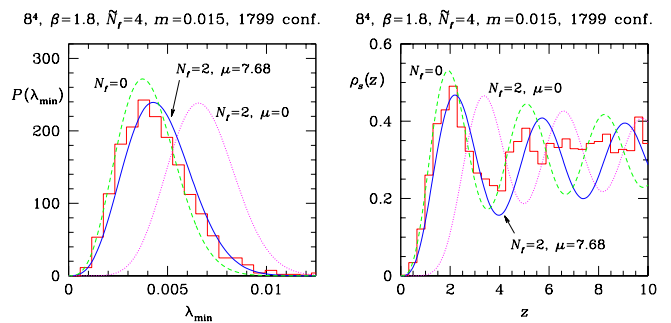}}
  \vspace*{3mm}
  \caption{Distribution of the smallest eigenvalue (left) and
    microscopic spectral density (right) of the lattice Dirac operator
    for the simulation parameters indicated above the figures.  The
    histograms represent the lattice data.  The full lines are the
    RMT-predictions for $N_f=2$ and $\mu=mV\Sigma=7.68$, the dashed
    lines those for the quenched approximation, and the dotted lines
    those for $N_f=2$ in the chiral limit, respectively.  Note that
    there is no free parameter involved.}
  \label{fig1}
\end{figure}

Our results for $\beta=1.8$ and $\tilde N_f=4$ flavors of mass
$ma=0.015$ are plotted in Fig.~\ref{fig1}.  Note that we have to
compare the data with the RMT-predictions for $N_f=2$ in
Eq.~(\ref{eq6}) for the following two reasons.  First, because of the
finite lattice constant the original ${\rm U}(4)\otimes{\rm U}(4)$
chiral symmetry of the (local) lattice action is broken down to ${\rm
  U}(1)\otimes{\rm U}(1)$.  This would imply to use $N_f=1$ in
Eq.~(\ref{eq6}).  Second, because of a global charge conjugation
symmetry which is special to SU(2) all eigenvalues are twofold
degenerate.  Equation~(\ref{eq6}) was derived for fermions in the
adjoint representation, and the eigenvalues in the fermion
determinants of Eq.~(\ref{eq6}) are assumed to be non-degenerate
\cite{Verb94a}.  Therefore, the number of flavors must be doubled.
Note that this is only an issue for SU(2).  For these two reasons, we
necessarily have to use $N_f=2$ in the random-matrix results.

After these comments, let us discuss Fig.~\ref{fig1}.  The presence of
the fermion determinants leads to a suppression of small eigenvalues.
The fact that the lattice data differ from the RMT-curves for the
quenched approximation ($N_f=0$, or $\mu=\infty$) means that the
dynamical quarks were light enough to affect the microscopic spectral
quantities.  However, the data are still closer to the RMT-curves for
$N_f=0$ than to the ones for the chiral limit ($\mu=0$).  This means
that the dynamical quarks with a rescaled mass of $\mu=7.68$ are still
quite heavy as far as the microscopic spectral properties are
concerned.  We will explore smaller masses in Fig.~\ref{fig2} below.
The lattice data agree reasonably well with the random-matrix
predictions for $N_f=2$ with the appropriate value of $\mu$.  The
systematic deviations and the question of the continuum limit will be
discussed below and in the conclusions.

For a larger number of flavors, the curves for the quenched
approximation and those for the chiral limit are farther apart from
each other so that the influence of massive dynamical quarks can be
seen more easily.  Therefore, we have also performed lattice
simulations with $\tilde N_f=8$ flavors, shown in Fig.~\ref{fig2}.  A
value of $\beta=1.3$ was chosen to ensure that we are in the broken
phase for all values of the bare quark mass \cite{Meye90}.  As
expected from the above discussion, we have to compare the lattice
data with the RMT-results for $N_f=4$ and the appropriate value of
$mV\Sigma$.  We see in Fig.~\ref{fig2} that, as the mass of the
dynamical quarks is lowered, the lattice data move away from the
quenched curves towards the RMT-curves computed in the chiral limit.
This is precisely what we expect.  Let us discuss the numbers.  Our
qualitative criterion was $\mu\sim{\cal O}(1)$.  For $\mu=8.74$ (the
top two figures in Fig.~\ref{fig2}), we are still rather close to the
quenched RMT-curves.  For $\mu=4.26$ (the bottom two figures in
Fig.~\ref{fig2}), we are about half-way in between the quenched
approximation and the chiral limit.  We can now say more
quantitatively that values of $\mu$ around 10 yield results which are
close to those obtained in the quenched approximation, whereas values
of $\mu$ smaller than 4 yield results which are closer to those
obtained in the chiral limit.  The mass scale on which dynamical
quarks affect the microscopic spectrum of the Dirac operator is thus
identified quantitatively.
\begin{figure}[-t]
  \centerline{\epsfig{figure=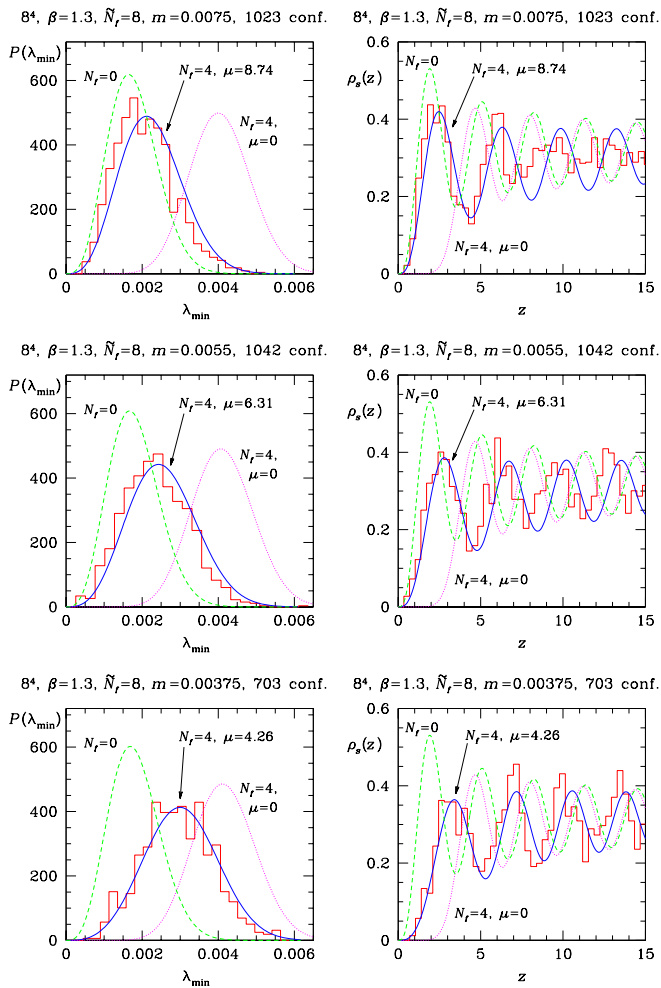}}
  \vspace*{3mm}
  \caption{Same as Fig.~\ref{fig1} but for different simulation
    parameters (indicated above the figures).} 
  \label{fig2}
\end{figure}

\begin{figure}
  \centerline{\epsfig{figure=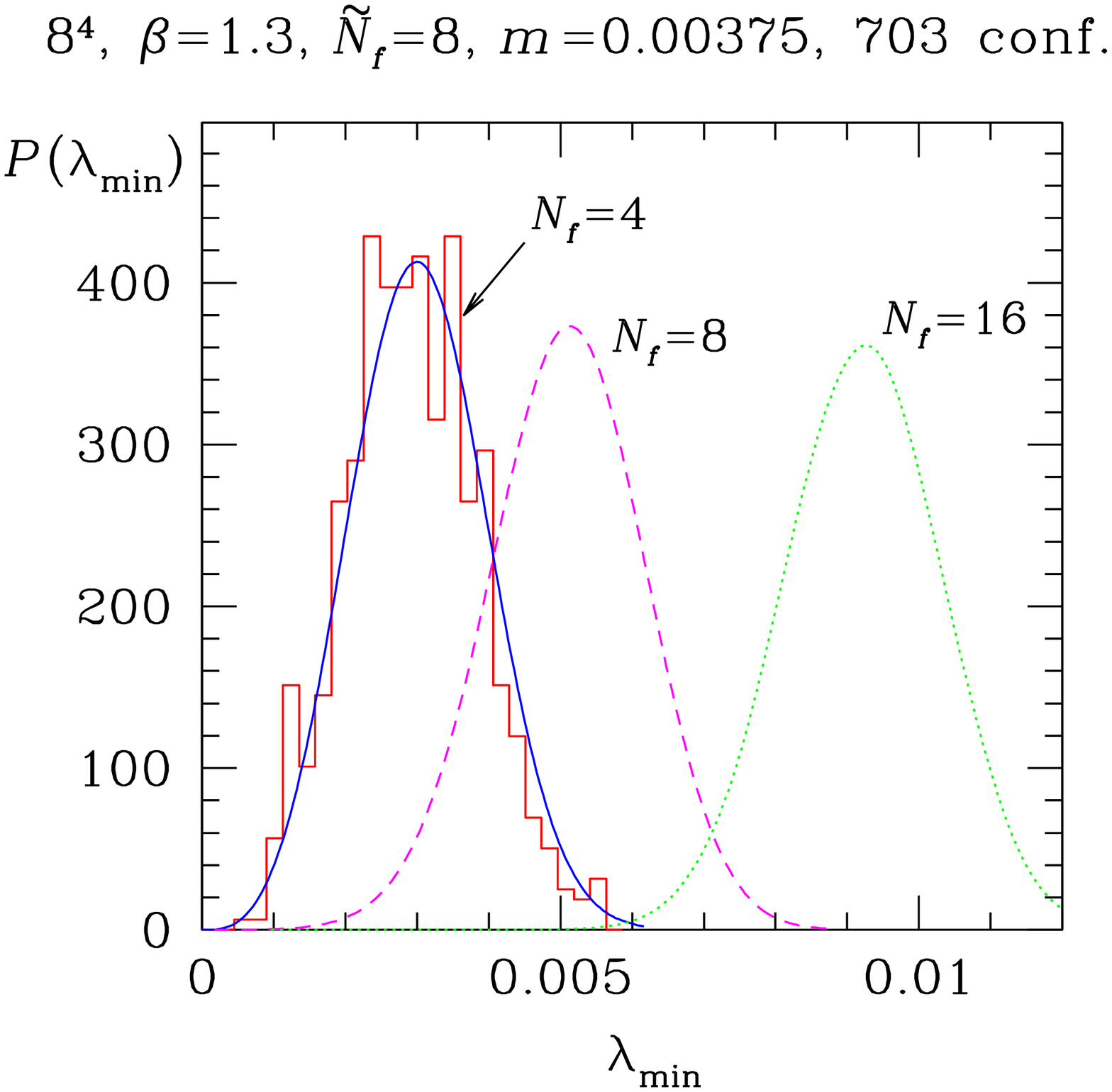,width=60mm}}
  \vspace*{3mm}
  \caption{Distribution of the smallest eigenvalue for $\tilde N_f=8$
    and $m=0.00375$ (histogram) compared with the RMT-pre\-di\-ctions
    for various numbers of flavors (indicated in the figure) and
    $\mu=4.26$.}
  \label{fig3}
\end{figure}

In Fig.~\ref{fig3}, we compare the data for $P(\lambda_{\rm min})$
computed for $\tilde N_f=8$ and the lightest quark mass, $m=0.00375$,
with the RMT-predictions for various numbers of flavors and the
appropriate value of $\mu$.  The fact that the data agree well with
the RMT-result for $N_f=4$ supports the two arguments made in
connection with Fig.~\ref{fig1}.  In the continuum limit, one would
expect agreement with the RMT-curve for $N_f=16$.  

The agreement between the various lattice data and the RMT-predictions
is not perfect.  There are some systematic deviations which we believe
to be largely due to the finite size of the lattice.  Similar
finite-size effects were observed in Ref.~\cite{Berb98}.  The present
exploratory study was restricted to a relatively small lattice size
since a large number of independent configurations is needed.  One
also observes from the figures that the agreement with RMT is better
for smaller values of $\mu$, as expected from the discussion in the
introduction.  While it would be desirable to perform a comprehensive
study for larger lattices and a number of different values of $\beta$,
$N_f$, and $m$, the agreement between lattice data and random-matrix
results seen in our figures is already very encouraging.  We point out
again that there is no free parameter involved.  Another possibility
would have been to fit the lattice data to the RMT-results by
adjusting the parameter $V\Sigma$.  This is an alternative way of
determining the chiral condensate which also seems to eliminate some
finite-size effects \cite{Berb97b}.  In this case, the agreement in
Figs.~\ref{fig1} and \ref{fig2} would have been better.  However, the
primary purpose of this work is the demonstration of parameter-free
agreement between lattice data and RMT-predictions.  This is the
reason why we determined $\Sigma$ from $\rho(0)$ and not from a fit to
RMT.

To conclude, we comment on the implications of our finding that
lattice data for the small eigenvalues of the Dirac operator are
described by chiral RMT also in the presence of dynamical quarks of
mass $\sim 1/(V\Sigma)$.  First of all, as far as the microscopic
spectral properties are concerned, the dynamical quarks are only
really dynamical if their masses are not much larger than
$1/(V\Sigma)$.  This provides information on the relevant mass scales
in unquenched lattice simulations.  Once the applicability of chiral
RMT to full QCD with massive dynamical quarks is firmly established,
one can address practical applications.  For quantities which are
sensitive to the small eigenvalues, the analytical RMT-results can
provide guidance for extrapolations to the chiral limit.  Also, as
demonstrated in Ref.~\cite{Berb97b}, extrapolations to the
thermodynamic limit are facilitated since the RMT-results are derived
in the limit $V\to\infty$.  Another interesting aspect which deserves
further investigation is the continuum limit where the original chiral
symmetry of the action is restored and agreement with the
RMT-results for the original number of flavors is expected.  This
might give an indication of how close to the continuum limit one
actually is.  However, much larger lattices are required to study such
a transition.  Moreover, we hope that the microscopic universality can
perhaps be used in the development of hybrid fermionic algorithms that
take into account the available analytical information.  It would also
be of great interest to extend our analysis to gauge group SU(3) where
analytical RMT-results are already known.  Finally, we hope to obtain
analytical results for the chSE with massive dynamical quarks in the
near future.

It is a pleasure to thank M. G\"ockeler, P. Rakow, A. Sch\"afer,
J.J.M. Verbaarschot, and H.A.  Weidenm\"uller for useful discussions.
We also thank the MPI f\"ur Kernphysik, Heidelberg, for hospitality
and support.  This work was supported in part by DFG grant We
655/11-2.  The numerical simulations were done on a CRAY T3E-900 at
the HLRS Stuttgart.

\end{document}